\documentclass{article}
\usepackage{ltwol2e}
\usepackage{epsfig}

\def\ifmath#1{\relax\ifmmode #1\else $#1$\fi}%
\newcommand {\ee}         {{e}^+{e}^-}
\newcommand {\ipb}   {\mbox{pb$^{-1}$}}
\newcommand{\bit}{\begin{itemize}}
\newcommand{\eit}{\end{itemize}}
\newcommand{\beq}{\begin{equation}}
\newcommand{\eeq}{\end{equation}}

\newcommand{\Ebeam}{\ifmath{E_{\rm{beam}}}}
\newcommand{\ECM}{\ifmath{E_{\rm{CM}}}}
\newcommand{\Epol}{\ifmath{E_{\rm{pol}}}}
\newcommand{\Enmr}{\ifmath{E_{\rm{NMR}}}}
\newcommand{\Bnmr}{\ifmath{B_{\rm{NMR}}}}
\newcommand{\Bfl}{\ifmath{B_{\rm{FL}}}}

\begin{document}

\title{EVALUATION OF THE LEP CENTRE-OF-MASS ENERGY\\
ABOVE THE W-PAIR PRODUCTION THRESHOLD}

\author{Michael D. Hildreth, \\for the
LEP Energy Working Group}

\address{CERN, CH1211, Geneva 23, Switzerland\\E-mail: mike.hildreth@cern.ch}   


\twocolumn[\maketitle\abstracts{Knowledge of the beam energy at LEP2 
is of primary importance to set the absolute energy scale
for the measurement of the W-boson mass.
The beam energy above 80~GeV is derived from continuous
measurements of the magnetic bending field by
16 NMR probes situated in a number of the LEP dipoles. 
The relationship between the fields measured
by the probes and the
beam energy is calibrated against 
precise measurements of the average 
beam energy between 41 and 55~GeV made using
the resonant depolarisation technique.
The linearity of the relationship is tested by
comparing the fields measured by the probes with the
total bending field measured by a flux loop.
Several further corrections are applied to derive
the the centre-of-mass energies
at each interaction point.
The beam energy 
has been determined
with a precision of 25~MeV for the data taken in 1997,
corresponding to a relative precision of $2.7\times 10^{-4}$.
Prospects for improvements are outlined. }]

\section{Introduction}

The large electron-positron (LEP) collider has been running
above the W-pair production threshold since 1996,
when the LEP2 programme began. 
The beam energy at LEP2 is used to set the absolute
energy scale for the measurement of the W-boson mass,
resulting in a relative uncertainty of 
$\Delta M_{\rm W} / M_{\rm W}  \approx
\Delta \Ebeam/ \Ebeam$.
The statistical uncertainty 
on the W mass, $M_{\rm W} \approx 80.4$~GeV
is expected to be around 25~MeV with the
full LEP2 data sample, which sets a target for a 
relative precision of around 10$^{-4}$ for the
beam energy to avoid this leading
to a significant contribution to the error. For a beam energy
of around 90~GeV, the aim is for a beam energy
uncertainty at the 10 to 15~MeV level.
This contrasts with
LEP1, where the Z mass was measured with an 
uncertainty  of about 2~MeV ($ 2 \times 10^{-5}$ relative),
corresponding to beam energy uncertainties of typically 1~MeV.

The centre-of-mass energy is derived by first determining the
average beam energy around the ring. In practice this involves
normalising the overall energy scale with respect to 
a precise reference, and in addition tracking time
variations of the beam energy.
Further corrections are
applied to obtain the $e^+$ and
$e^-$ beam energies at the four interaction
points, and the centre-of-mass energy in the 
$\ee$ collisions.

At LEP1, the average beam energy 
was measured directly at the physics operating energy
with a precision of better than 1~MeV by 
resonant depolarisation (RD)\cite{ref:zpaper}. 
The spin tune, $\nu$, determined by RD, and the 
average beam energy, are both
proportional to the total integrated
vertical magnetic field, $B$, along the beam trajectory, $\ell$:
\begin{equation}
\nu \propto \Ebeam \propto \oint B \cdot \rm{d} \ell .
\end{equation}
Unfortunately, the RD technique can not be used at LEP2. 
A sufficient level of transverse polarisation was not achieved
above a beam energy of 55~GeV in 1997, due to the 
sharp increase of depolarising effects with beam energy.
Instead, the beam energy is determined from an
estimate of the field integral derived from continuous
magnetic measurements by 16 NMR probes situated in some 
of the LEP dipoles. 
The relation between their readings and the beam energy can be 
precisely calibrated in the beam energy range 41 to 55~GeV.

This relation is assumed to be linear, and to be valid at
physics energies. The linearity can only be tested over
a limited range with the RD data themselves.
A second comparison of the NMR readings with
the field integral, which is proportional to the beam energy,
is made using the flux loop. 
The flux loop experiments provide
a measurement of almost the complete field integral (97\%).
The local bending fields measured by the
NMR probes are read out
over the full range from
RD to physics energies, allowing an independent constraint
on the linearity of the relation between the probe fields
and the total bending field.

The use of the NMR probes to transport the precise energy scale
determined by RD to the physics operating energy is the 
main novelty of this analysis. The systematic errors on the
NMR calibration are evaluated from the reproducibility of
different experiments, and the variations from probe to probe.
The dominant uncertainty comes from the flux-loop test.

In addition, the other effects that were
extensively studied at LEP1~\cite{ref:zpaper} which can modify the beam energy,
such as earth tides, leakage currents from trains, {\it etc.}, must be
taken into account when comparing the NMR measurements
with the RD beam energies, and in deriving the centre-of-mass
energy of collisions as a function of time.

The exact beam energy along the ring differs from the average
because of the loss of energy by synchrotron radiation in the
arcs and the gain of energy in the RF accelerating sections,
which must be taken into account in determining the centre-of-mass
energy at each IP.
The centre-of-mass energy at each collision point can also be
different from the sum of the beam energies due to 
the interplay of collision offsets and dispersion. 

A more detailed version of this contribution can be found in
\cite{ref:LEP2paper}.

\section{Data samples}
\label{sec:data}

\subsection{Luminosity delivered by LEP2}

LEP has delivered about 10\ipb\ at each of two centre-of-mass
energies, 161 and 172~GeV, in 1996, and over 50\ipb\ at a 
centre-of-mass energy of around 183~GeV in 1997. Combining
the data from all four LEP experiments, these data give
a measurement of the W mass with a precision of 
about 90~MeV~\cite{ref:thomson}. This paper emphasises
the 1997 energy analysis.

\subsection{Polarisation measurements}

\begin{table}[htb]
\begin{center}
\begin{tabular}{l|c|c|c|c|c|c}
\hline
 & &  \multicolumn{4}{|c|}{Beam Energy [GeV]} & \\
Date     & Fill   & 41 & 44 & 50 & 55 & Optics \\
\hline
19/08/96 & 3599  &        &     & yes &    & 90/60 \\
31/10/96 & 3702  &        & yes &     &    & 90/60 \\
03/11/96 & 3719  &        & yes & yes &    & 90/60 \\
\hline
17/08/97 & 4000  &        & yes &             &    & 90/60\\
06/09/97 & 4121  &        & yes & yes &         & 60/60\\
30/09/97 & 4237  &        & yes & yes &         & 60/60\\
02/10/97 & 4242  & yes & yes & yes & yes   & 60/60\\
10/10/97 & 4274  &        & yes &          &       & 90/60\\
11/10/97 & 4279  & yes & yes & yes & yes   & 60/60\\
29/10/97 & 4372  & yes & yes &        &         & 60/60 \\
\hline
\end{tabular}
\end{center}
\caption{Fills with successful polarisation measurements in 1996
and 1997. Successful measurements are marked ``yes''.}
\label{tab:pol}
\end{table}

The successful RD experiments in 1996 and 1997 
are listed in table~\ref{tab:pol}.
To reduce uncertainties from fill-to-fill variations,
an effort was made to measure as many 
beam energies as possible with RD during
the same LEP fill.
In 1997, improvments in the orbit quality and reducing depolarising
effects resulted in
5 fills with more than one energy point, and
2 fills with 4 energy points, which allow a check
of the linearity assumption.

\subsection{Magnetic measurements}

The LEP dipole fields are monitored in two ways. There are 16 
NMR probes positioned inside some selected main bend dipoles. 
The probes measure the local 
magnetic field
with a precision of around $10^{-6}$, and can be read out
continually, but each probe only
samples the field in a small region of one out of 3200 dipoles.
The probes are read out during normal physics running, and also
during RD and flux-loop measurements.

In contrast to the NMR probes, the flux loop measures 
96.5\% of the total bending field of LEP,
including 98\% of the main bend dipole
field.

The flux loop measures the change in field during
a dedicated demagnetisation cycle, outside physics running. 
This possibility to cross-calibrate the field measured
by the flux loop and by the NMR probes is crucial to the analysis.

\section{The beam energy model}
\label{sec:model}

The LEP beam energy is calculated as a function of time 
according to the following formula:
\begin{eqnarray}
\Ebeam (t) & = &  (E_{\rm{initial}} + \Delta E_{\rm{dipole}}(t) )
 \label{eq:model}\\
& &
 \cdot  (1+C_{\rm{tide}} (t))\cdot  (1+C_{\rm{orbit}}) 
\cdot (1 + C_{\rm{RF}}(t)) \nonumber \\
& &
 \cdot  (1+C_{\rm{h.corr.}} (t))\cdot  (1+C_{\rm{QFQD}} (t)) \nonumber  . 
\end{eqnarray}
The first term, $E_{\rm{initial}}$, is the energy at
the start of the fill derived from the estimate of
the integral dipole bending field at the end of the
LEP ramp up to operating energy,
and $\Delta E_{\rm{dipole}}(t)$ is 
the shift in energy caused by changes in the bending dipole
fields during a fill. Both are averages over the energies 
predicted by each functioning NMR probe. 
This is a simplification over the treatment at LEP1, 
since the energy is taken from the average of the measured
magnetic fields, rather than a model of the field evolution.

The remaining terms correct for other contributions to the
integral bending field.
They are discussed
in more detail 
in references~\cite{ref:zpaper,ref:LEP2paper}.
The effect of earth tides which
move the quadrupole magnets with respect
to the fixed length
beam orbit is accounted for by $C_{\rm{tide}} (t)$.
Distortions of the ring geometry on a longer time scale are 
corrected for by the measured average horizontal orbit displacement,
$C_{\rm{orbit}}$
which is evaluated for each LEP fill.
Regular changes in the RF frequency away from the nominal central
frequency are made to optimise the luminosity, and are accounted for by
$C_{\rm{RF}}(t)$.
The term $C_{\rm{h.corr.}} (t)$ accounts for 
changes in the beam steering by horizontal orbit corrector magnets,
and the term $C_{\rm{QFQD}} (t)$ takes into account the
effects of stray fields due to 
different excitation currents 
in the focussing and defocussing quadrupoles in the LEP lattice.

\section{Calibration of NMR probes}
\label{sec:nmrcal}

\subsection{Calibration of NMR probes with RD measurements}
\label{sec:nmrpol}

The magnetic fields $\Bnmr^i$ measured by each NMR $i=1,16$,
are converted into an equivalent beam energy. The 
relation is assumed to be linear, of the 
form
\beq
 \Enmr^i = a^i + b^i \Bnmr^i.
\eeq
In general, the beam energy is expected to be
proportional to the integral bending field. The offset
allows remnant fields or stray constant fields such as the
earth's field to be absorbed.
The two parameters for each probe are
determined by a fit to the energies measured
by resonant depolarisation. The NMR probes give
an estimate of the dipole contribution to the integral
bending field, so the energy measured by RD must
be corrected using equation~\ref{eq:model} for
the effects of coherent quadrupole motion etc.

\begin{figure}[htb]
\centering

\mbox{\epsfxsize=0.45\textwidth
\epsfbox{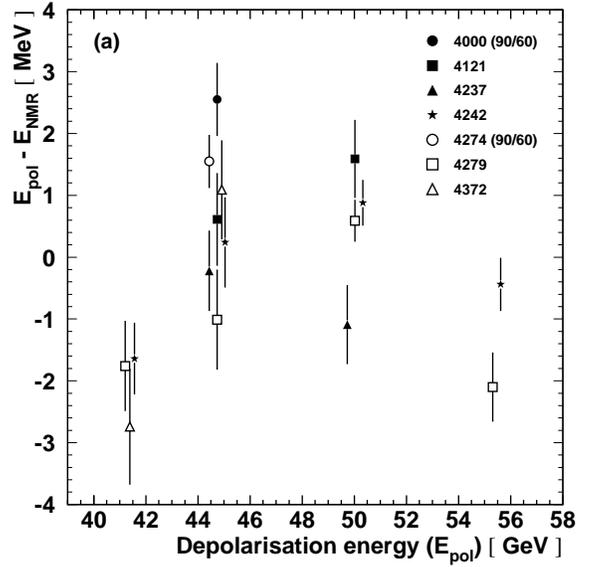} }

\caption{Residuals of the fit comparing RD energies to the energies
predicted by the model for a simultaneous fit to all fills}
\label{fig:allresid}
\end{figure}

The residuals, $\Epol - \Enmr^i$, are examined for each NMR.
The residuals evolve with beam energy in a 
different way for different probes, 
but for a particular probe
this behaviour is reproduced from fill to fill.
The residuals averaged over NMR probes at each polarisation point 
are shown in figure~\ref{fig:allresid},
in which the errors are displayed as the rms/$\sqrt{N}$, where $N\le 16$
is the number of NMR probes functioning for the measurement.
This figure shows the average residuals with respect to the simultaneous fit
to all polarisation fills in 1997, which was used to calibrate the
NMR probes. The residuals show a reproducible small
but statistically significant deviation from zero, with the 
45 and 50~GeV points being a few MeV higher than those at
40 and 55~GeV. 

\subsection{Test of NMR calibration using the flux loop}

\begin{figure}[b]
\centering
\mbox{\epsfxsize=0.45\textwidth
\epsfbox{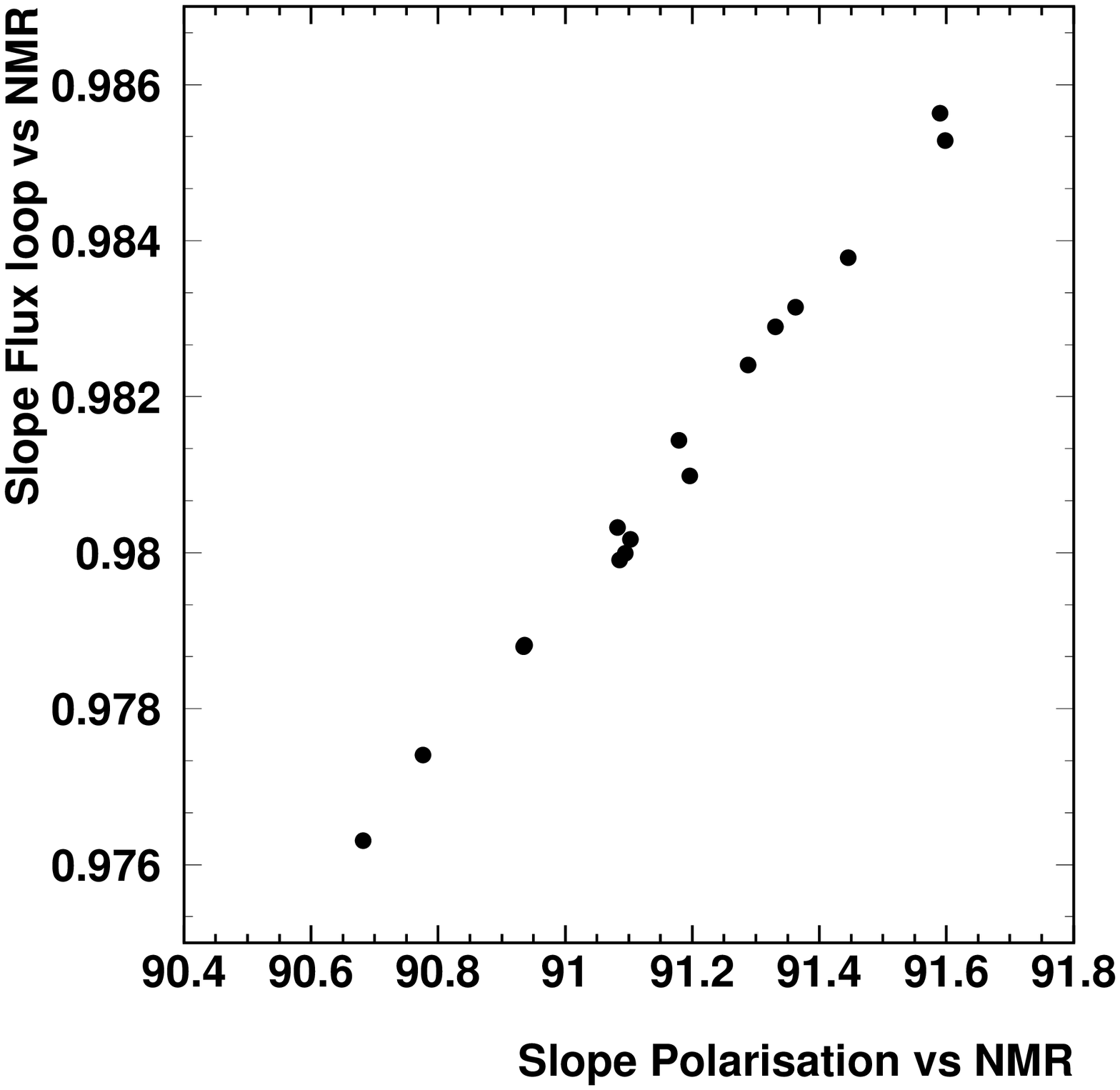} }
\caption{The slopes of 
equations~\ref{eq:correlab} and \ref{eq:correlcd}, showing good correlation.}
\label{fig:correl}
\end{figure}

The total change of field measured by the flux loop can also
be fit to a straight line as a function of NMR magnetic
field for each probe:
\beq
 \Bfl  =  c^i + d^i \Bnmr^i
\eeq
From octant-to-octant and magnet-to-magnet the probes show
different behaviour; the slopes are not constant with energy. 
However, the behaviour is reproduced from fill-to-fill. 

As would be expected
if the RD measured energies, \Epol, are proportional to the 
total bending field measured by the flux loop, \Bfl, there is 
a strong correlation between the parameters
of the polarisation and the flux loop fits, 
when the NMR/flux-loop comparison is restricted
to the range corresponding to 40--55~GeV:
\begin{eqnarray}
\label{eq:correlab}
 \Epol  =  a^i + b^i \Bnmr^i &&\mbox{and}\\
\label{eq:correlcd}
 \Bfl  =  c'^i + d'^i \Bnmr^i && \mbox{fit restricted to 40--55 GeV}.
\end{eqnarray}

Having established this correlation, which is shown in figure~\ref{fig:correl},
the flux loop can now be used
to test the extrapolation method.
From the fit in the 40--55~GeV region, the NMR probes can be used to 
predict $\Bfl$ at physics energy, and this 
prediction can be compared with the flux loop measured \Bfl.
Any difference between the prediction and the direct measurement
of the field
can then be interpreted as a possible bias in the predicted energy.

\begin{figure}[htb]
\begin{center}
\mbox{\epsfxsize=0.45\textwidth
\epsfbox{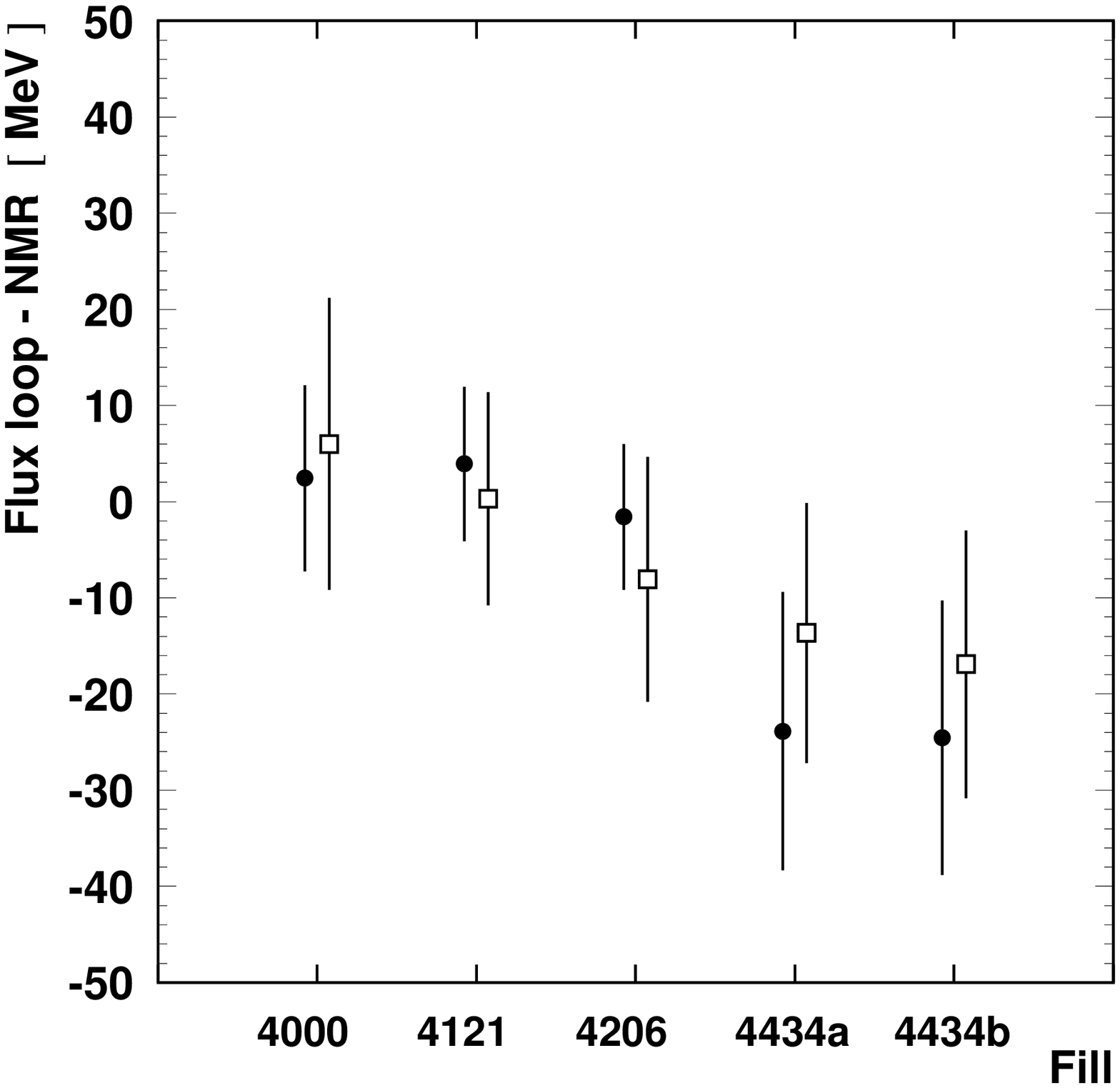} }
\end{center}
\caption{The difference in energy between flux loop and NMR probes
at the equivalent of physics energy 
for each flux loop measurement. The solid points are for all
working probes for each fill, while the empty points are
for the common set of probes working for all fills.}
\label{fig:fluxb}
\end{figure}

The deviations at
physics energy for all five 
flux loop measurements are shown in figure~\ref{fig:fluxb}. 
The bias
at physics energy is up to $20$~MeV, with an rms
over the probes of 30--40~MeV, corresponding to an uncertainty
of 10--20~MeV depending on the number of working probes.
The bias tends to decrease during the year.
No cause for this change has been identified so far. 

The tests with the flux loop are not used to correct the NMR 
calibration from polarisation data, but allow an independent estimate
of the precision of the method. The difference between the NMR and
flux-loop measured field increase 
is taken to assign
a systematic uncertainty of 20~MeV. This covers the maximum difference
seen during the year. 

\section{Evaluation of the centre-of-mass energy at each IP}
\label{sec:ipcor}

As at LEP1, corrections to the centre-of-mass energy arise from the non-uniformity
of the RF power distribution around LEP and from possible offsets of the beam
centroids during collisions in the presence of opposite-sign vertical dispersion\cite{ref:zpaper}.

\subsection{Corrections from the RF System}
Since the beam energy loss due to synchrotron radiation is
proportional to $E_{\rm {beam} }^4$, operation of LEP2 requires a
large amount of RF acceleration to maintain stable beam orbits.  
This implies that the energy variation in the beams (the ``sawtooth'') as they circulate
around LEP is quite large, 
which increases the sensitivity of the centre-of-mass
energy to non-uniformities in the energy loss arising from differences in the local
magnetic bend field, machine imperfections, etc.

The average corrections for the 1996 and 1997 running are about 20 MeV in IPs 2 and 6,
and -5 MeV at IPs 4 and 8.

As at LEP1, the errors on the energy corrections are evaluated by a comparison
of those quantities (the synchrotron tune $Q_s$, the orbit sawtooth, and the longitudinal position of the
interaction point)
which can be calculated in the RF model and
can be measured in LEP.  In addition, uncertainties from the inputs to
the model, such as the misalignments of the RF cavities and the
effects of imperfections in the LEP lattice, must also be considered.
This results in a systematic error of 4 MeV per beam.

\subsection{Opposite sign vertical dispersion}

In the bunch train configuration, beam offsets at the collision point
can cause a shift in the centre-of-mass energy due to opposite sign
dispersion~\cite{ref:zpaper}. 
The change in energy is evaluated from the calculated
dispersion and the measured beam offsets from beam-beam deflections scans.

Beam offsets were
controlled to within a few microns by beam-beam deflection scans. 
The resulting luminosity weighted correction to the centre-of-mass 
energies are typically 1 to 2~MeV, with an error of about 2~MeV.
No corrections have been applied 
for this effect, and an uncertainty of 2~MeV has been assigned.

\section{Summary of systematic uncertainties}
\label{sec:syst}

\begin{table}[htb]
\begin{center}
\begin{tabular}{l|c}
\hline
Source                                     & Error [MeV]\\
\hline
 Extrapolation from NMR--polarisation:      &    \\
\hspace*{0.2cm} NMR rms/sqrt(N) at physics energy          & 10 \\
\hspace*{0.2cm} Different \Epol\ fills                     & 5  \\
\hline
 Flux-loop test of extrapolation:           &               \\
\hspace*{0.2cm} NMR flux-loop diff. at physics energy& 20 \\
\hspace*{0.2cm}  Field not measured by flux loop           & 5  \\
\hline
Polarisation systematic                    & 1 \\
$\ee$ energy difference                    & 2 \\
Optics difference                          & 4 \\
\hline
Corrector effects                          & 3  \\
Tide                                       & 1  \\
Initial dipole energy                      & 2  \\
Dipole rise modelling                      & 1  \\
\hline
IP specific corrections ($\delta \ECM/2$):&    \\
\hspace*{0.2cm} RF model                                   & 4  \\
\hspace*{0.2cm} Dispersion                                 & 2  \\
\hline
Total                                      & 25 \\
\hline
\end{tabular}
\end{center}
\caption{Summary of contributions to the beam energy 
uncertainty.}
\label{tab:error}
\end{table}

The contributions from each source of uncertainty
described above are summarised in table~\ref{tab:error}.
The first groups describe the uncertainty in the normalisation
derived from NMR-polarisation comparisons, NMR-flux-loop tests and
the part of the bending field not measured by the flux-loop. These
extrapolation uncertainties dominate the analysis.
The following errors concern the 
polarisation measurement, specifically its intrinsic 
precision (which is less than 1~MeV),
the possible difference in energy between electrons
and positrons, and the difference between optics.
None of the additional uncertainties  from 
time variations in a fill, and IP specific  
corrections contribute an uncertainty greater than 5~MeV. 

\subsection{Uncertainty for data taken in 1996}

The analysis of the 1996 data was 
largely based on a single fill with RD measurements 
at 45 and 50~GeV. The apparent consistency
of the flux-loop and NMR data compared to RD data 
was about 2~MeV over this 5~GeV interval, 
i.e.\ a relative error $4 \times 10^{-4}$, which using
a naive linear extrapolation 
would give an uncertainty of 13.5 (15)~MeV at 81.5 (86)~GeV. 
These errors were inflated to 27 (30)~MeV before the 1997 data were 
available, since there was no test of reproducibility from fill to fill, 
there was no check of the non-linearity possible from a fill
with two energy points, and the field outside the flux loop had
not been studied.
It can be assumed that
the 25~MeV uncertainty of 1997 data is common to the 1996 data.

\section{Conclusions and outlook}
\label{sec:conclude}

The method of energy calibration by magnetic extrapolation of
resonant depolarisation measurements at lower energy has made
substantial progress with the 1997 data. The success
in establishing polarisation above the Z has allowed a 
robust application of the method, and 
the mutual consistency of the resonant depolarisation, 
NMR and flux-loop  data
has been established at the 
20~MeV level at physics energy, with a total systematic
uncertainty in the beam energy of 25~MeV. 
The precision is limited by the understanding of the
NMR/flux-loop comparison. 

As LEP accumulates more high energy data, the experiments
themselves will be able to provide a cross-check on the centre-of-mass
energy by effectively measuring the energy of the emitted photon in
events of the type $e^+ e^- \to {\rm Z} \gamma \to f \bar f \gamma$,
where the Z is on-shell.  This can be done using a kinematic fit
of the outgoing fermion directions and the precisely determined Z-mass
from LEP1.  The ALEPH collaboration have shown\cite{ref:aleph_ebeam}
the first attempt to make this measurement in the $q\bar q \gamma$ channel,
where they achieve a precision of 
$\delta\Ebeam = \pm 0.110 ({\rm stat}) \pm 0.53 ({\rm syst})$.  With 500 pb$^{-1}$
per experiment, the statistical precision on this channel should approach 15~MeV.
Careful evaluation of systematic errors will determine the usefulness of this approach.

In future, a new apparatus will be available for measuring
the beam energy.
The LEP Spectrometer Project~\cite{ref:massimo}
will measure the bend angle of the LEP beams using 
beam pick ups with new electronics to measure the position
to the order of a micron precision as they enter or leave
a special dipole in the LEP lattice whose bending field
has been surveyed with high precision. A first phase
of the spectrometer is already in place for the 1998 running,
with the aim of checking the mechanical and thermal stability
of the position measurement. In 1999, the new magnet will
be installed, and the aim is to use this new, independent
method to measure the beam energy to 10~MeV at high energy.
It should be possible to propagate any improvement in
the beam energy determination back to previous years
by correcting the extrapolation and correspondingly
reducing the uncertainty.

\end{document}